# Molecular evolution and gene function

Marc Robinson-Rechavi[1,2]

1: Department of Ecology and Evolution, University of Lausanne, 1015 Lausanne, Switzerland

2: Swiss Institute of Bioinformatics, 1015 Lausanne, Switzerland

## 1. Abstract

One of the basic questions of phylogenomics is how gene function evolves, whether among species or inside gene families. In this chapter, we provide a brief overview of the problems associated with defining gene function in a manner which allows comparisons which are both large scale and evolutionarily relevant. The main source of functional data, despite its limitations, is transcriptomics. Functional data provides information on evolutionary mechanisms primarily by showing which functional classes of genes evolve under stronger or weaker purifying or adaptive selection, and on which classes of mutations (e.g., substitutions or duplications). However, the example of the "ortholog conjecture" shows that we are still not at a point where we can confidently study phylogenomically the evolution of gene function at a precise scale.

## 2. The problem with "function"

Molecular evolution interacts with gene function in two fundamental ways. First, different gene families will evolve differently according to their function, e.g. they are under different selection pressures on their protein sequence or on their diversification by gene duplication (see below). Second, gene function itself evolves. Both of these assertions are quite obvious in their generality. Problems arise when we try to characterize more specific patterns, and to test more specific hypotheses. While no aspect of phylogenomics is without its difficulties, this is a particularly vexing one: what is gene function?

Two distinctions are fundamental to the study of function. First, between healthy and pathological function, i.e. what the gene does when it is present and functional, versus what is disrupted when the gene is absent or somehow not functioning properly.

The latter includes most medical genetics observations, as well as Knock-Out/Knock-Down phenotypes. Second, we need to distinguish between selected effect and causal role. This second distinction has been abundantly discussed following the publication of ENCODE 2012 (Pennisi 2012; The ENCODE Project Consortium 2012; Doolittle 2013; Eddy 2013; Graur et al. 2013; Germain et al. 2014; Graur et al. 2015). ENCODE is a large collaborative project to "build a comprehensive parts list of functional elements in the human genome", based on systematic biochemical assays, such as RNA-seq or ChIP-seq, in different cell types. The observation that ≈80% of the human genome had some type of biochemical activity in some cell type led to statements that all that DNA was functional (Pennisi 2012; The ENCODE Project Consortium 2012). The questions of function and of evolution are tightly linked in biology because it is natural selection which explains the functional adaptation of organisms and their parts. The function of the lungs is to breath, i.e. to exchange oxygen and $CO_2$ between the organism and the air. This comes neither from intention of the lungs nor of the organism, but because ancestors of some vertebrates which were better at exchanging oxygen and $CO_2$ with the air had better survival and reproductive success. Thus it has been proposed that function be defined as that which a structure was selected to do. This is the "selected-effect definition of function" (Doolittle et al. 2014). The lungs were selected to exchange gases, not to develop cancers or take space in the thoracic cage, although they also do these things. An alternative definition of function, the "causal role" definition, does not appeal to evolutionary history, and could in fact include such features as the lungs taking space, or the nose supporting sunglasses (Doolittle et al. 2014). The same questions and definitions apply to all levels of biological organization, including genes.

In the aftermath of ENCODE, much of the focus has been on classifying DNA sequences as "functional" or not. This question is more directly relevant to genome annotation. For this chapter, we will mostly focus on protein coding genes, for which we have strong a priori reasons to expect that they are indeed functional. One simple line of evidence is that genes which are sufficiently conserved among species to undertake phylogenomics studies are most probably conserved by purifying selection, and thus functional. But to understand the role of function in molecular evolution beyond the generality that functional sequences are more conserved, we need to focus on classifying their specific functions.

One way to classify specific gene functions is to collect assertions and evidence from the published biological literature (Thomas 2017). The largest undertaking in this sense is the Gene Ontology consortium (Box 1). The Gene Ontology describes the selected effect function of gene products, whether they are proteins or functional RNAs. Thus it notably does not describe pathological roles, which are typically causal role functions.

> The Gene Ontology is composed of three ontologies, which describe different aspects of gene function (Ashburner et al. 2000; Dessimoz and Skunca 2016). Briefly, the Cellular Component ontology describes where in or out of a cell the gene product is found; the Molecular Function ontology describes the activity of the gene product, potentially as part of a protein or RNA complex; the Biological Process ontology describes the result of the organismal program in which the gene product acts. As can be readily seen, the latter is more complex than the other two. The Molecular Function can be thought of as "what does the gene product do in a test tube?", while the Biological Process can be thought of as "what does the gene product do within the organism?". Being ontologies, all three include not only standard terms and definitions, but also relations between the terms. These relations form a directed graph, meaning that (i) there is a direction to the relations, for example "steroid binding" is_a "lipid binding" but not the inverse, and (ii) terms can have both several children and several parents, for example "steroid binding" not only is_a "lipid binding" but also is_a "organic cyclic compound binding" and has_input from "steroid" (parents in the graph), while it has ten children, including "steroid hormone binding" and "vitamin D binding". This graph includes very general terms, such as "binding" or "catalytic activity", and very specific terms, such as "17alpha-hydroxyprogesterone binding" or "estrogen response element binding".
>
> The annotation of genes with the Gene Ontology consists in associating each gene with as many Gene Ontology terms as necessary, which describe the known function of the product(s) of this gene. Association can be based on (i) evidence from hypothesis-driven, small-scale, published studies, which provide the closest to selected effect function; (ii) large scale hypothesis-free experiments (such as ENCODE), which provide "candidate functions" (Thomas 2017), closer to the causal role functional definition; or (iii) electronic inference, whether simply by "best Blast hit" or more advanced domain modelling or text mining.

**Box 1: The Gene Ontology**

From a phylogenomic perspective, the properties of the Gene Ontology and its annotations have important consequences. These annotations can only ever capture knowledge at a given point in time, and they capture it from a disparate collection of studies with differing aims and methods. Thus even genes with evolutionarily conserved functions will often have different annotations, because of different experiments (e.g., Altenhoff et al. 2012; Chen and Zhang 2012) (Table 1). Moreover many genes are never or very rarely the object of targeted experimental studies (Sinha et al. 2018).

| Evidence gene A | Evidence gene A' | Apparent conclusion | Relevance |
| --- | --- | --- | --- |
| Experiment X: function x | Homology transfer: function x | Conserved function | No: circular reasoning |
| Experiment X: function x | Experiment Y: function y | Different function | No: experiments cannot be compared |
| Experiment X: function x | Experiment X: function x | Conserved function | Yes: evolutionary conservation |
| Experiment X: function x | Experiment X: function x' | Different function | Yes: evolutionary change |

Table 1: Evidence for function of homologous genes and evolutionary relevance. A and A' are homologous genes.

These limitations are not specific of the Gene Ontology, but will affect any effort to capture gene function from the abundance of precise but heterogeneous experimental data. For example, Enzyme Classification (E.C.) numbers (McDonald and Tipton 2014) have been used to investigate functional evolution, but E.C. numbers are mostly associated to gene products by homology, at the gene or the domain level, thus creating pseudo-evolutionary patterns in the data. If all proteins with homology to a given enzyme obtain a certain E.C. number, then that function will appear conserved, whether it is or not (Table 1). In the GO, the evidence used for assertions of functional annotation are available in a standard code (Giglio et al. 2018), which allows to distinguish conservation of function between homologs with experimental evidence from patterns due to functional annotation transfer between homologs.

Directly comparing the phenotypes associated to genes is even more complicated by the differences among experiments and species (Box 2). A few studies have shown promise in that phenotypes can effectively be compared between distant species (McGary et al. 2010; Kachroo et al. 2015), but the complexity of phenotypes still limits applications such as comparing subtle changes between orthologs or paralogs (see Fernández et al. 2019 for definitions), or relating functional change to protein evolutionary rates.

> Within the selected-effect definition of function, an ideal measure of function would be to relate genes to organismal level phenotypes. But to use them in phylogenomic studies, we need to define and measure phenotypes in a way that is systematic and robust enough.
>    One basic measure of phenotype impact is essentiality: is loss of a gene lethal to the organism (often extended in sexual organisms to include sterility) (Hurst and Smith 1999; He and Zhang 2006; Liao and Zhang 2007; Makino et al. 2009)?

> While this seems straightforward, the same gene loss can be lethal or not depending on growth conditions (Ooi et al. 2006) or genetic background (Ayadi et al. 2012). This limits the evolutionary interpretation of such results, since natural selection has been acting on genes in a variety of backgrounds and environments.
>
> In unicellular cultivated organisms, such as many bacteria or yeasts, one standardised measure of phenotype for comparisons among paralogs or strains is growth rate in a controlled environment (Hillenmeyer et al. 2008). One positive aspect of such measures is that they are probably closely related to fitness, but on the other hand, they only convey a very unspecific characterization of gene function. To study phenotypes beyond essentiality at a genomic scale between species, they need to be encoded in a standard manner. One promising solution is to develop inter-species phenotype ontologies (Mungall et al. 2010; Robinson et al. 2014; Mungall et al. 2017), but this approach is still limited by the difficulties of annotating phenotypes in different species. A recent study measured growth phenotypes in 32 bacterial species over different conditions (Price et al. 2018). This still only covers a small part of the genes of these species, but it shows promise in the possibility of scaling up to full phylogenomic studies. However, this approach remains restricted to easily cultivated microorganisms.
>
> Finally, two caveats affect almost all measures of phenotype from gene Knock-Out experiments. First, the conditions under which natural selection has acted are expected to be very different from the typical laboratory settings (e.g., Ruff et al. 2015). Secondly, "knocking out" a gene can be done in different ways (complete or partial, conditional or not), and it is not obvious which of these correspond to mutations which could occur in nature and be subject to natural selection. For example comparing phenotypes of essentiality between human and mouse means comparing diverse experimental designs to diverse spontaneous mutations (Liao and Zhang 2008), or using essentiality in human cell culture.

**Box 2: Phenotypes and function**

An alternative approach to investigate specific gene function is to use genome-wide experiments. While such data have been criticized for biasing GO annotations towards the types of function that can thus be investigated (Schnoes et al. 2013), they can provide comparable functional information across genes and species. Transcriptomics is particularly interesting because techniques are becoming relatively cheap and straightforward to apply to different species, conditions, or individuals, thus providing a direct link between gene activity and evolution. Yet there are also limitations of these data. Gene expression does not provide information on most aspects of gene function. Transcriptomics informs on (i) where and when a gene is expressed, (ii) how highly it is expressed, and (iii) which genes are co-expressed, but gives little information about which components of the phenotype are involved. On the other hand, transcriptomics provides a direct link between phylogenomics and Evo-Devo, where expression patterns are the main form of evidence. From a phylogenomic perspective, while it is relatively straightforward to compare gene expression results between paralogs within a species, comparisons between species are more complicated (discussed in Roux et al. 2015). Indeed, the direct comparison of expression levels is complicated by batch effects (Gilad and Mizrahi-Man 2015), different organisms being often studied independently. On the other hand, transforming continuous expression values into "expressed" versus "not expressed", which allows comparison between different species and provides a link to Evo-Devo reasoning, loses much of the information from transcriptome data.

Correlations of expression levels in different conditions (e.g., different organs) are also problematic (Pereira et al. 2009; Piasecka, Robinson-Rechavi, et al. 2012). Some of these problems have been evaded by defining qualitative variables summarizing patterns of gene expression, such as tissue specificity, which reflects function while being robust to differences in methods and sampling (Kryuchkova-Mostacci and Robinson-Rechavi 2016; Kryuchkova-Mostacci and Robinson-Rechavi 2017). An additional complexity of using gene expression in phylogenomics is that samples must be comparable (discussed in Roux et al. 2015). In practice, different organs, developmental stages, sexes, or abiotic conditions can be sampled, and homology or even similarity are not always clear. Even inside one species, for instance when comparing paralogs, care must be taken to distinguish variation in expression across tissues or developmental sequences from changes between experimental, abiotic conditions.

Assuming that, despite these many caveats, functional annotation has been achieved in a large enough set of species, one can think about studying the evolution of gene function. Ideally, we would like to know when function changed, and whether the changes were driven by selection or drift. The main approach to this question is based on Ornstein-Uhlenbeck models, which are notably used in the phylogenetic study of gene expression (Bedford and Hartl 2009). Briefly, a Brownian model of gene expression change is contrasted to models with different optima in different lineages; if there is significant support for different optima, this can be taken as evidence for changes in gene function. While the principle is very attractive, the limited data that we still have leads to issues of lack of power or of over-fitting (e.g. Ho and Ané 2014; Cooper et al. 2016), and there are problems with phylogenetic studies of expression when species sampling is small (Dunn et al. 2013). Finally, summarizing the expression of many genes in modules is also attractive because of its relevance to the way genes are expected to function as modules in relation to biological processes. These modules can be computed per species, before evolutionary computations (e.g. Piasecka et al. 2013), or computed across species, allowing to detect conserved expression patterns (e.g. Brawand et al. 2011). The clustering itself can also contain information on gene evolution, for example with transcriptomes of eyes of cave-dwelling and surface crayfish clustering by eye function and not according to the phylogenetic relationships of the species (Stern and Crandall 2018). These aspects are developed further in section 4 below.

## 3. Gene families with different functions evolve differently

Gene function and evolution can interact in two ways: genes with different functions evolve differently, and the function itself evolves. The first aspect is easier to study, as it is less dependent on the detailed specifics of functional annotation. On the other hand, causality can be difficult to determine, as many features of gene function and evolution are correlated. We will present here some of the main trends, keeping in mind that this is a rapidly changing domain.

### 3.1. Gene expression and function determine protein evolutionary rates

The sequence of different proteins evolves at very different rates, over at least three orders of magnitude. Efforts to understand the reasons of this variation have been called a "quest for the universals of protein evolution" (Rocha 2006). The most intuitive explanation for these differences is that proteins that are more essential to the organism evolve slower, because of stronger negative selection (selection against change). But studies of the statistical determinants of protein evolutionary rates have shown that reality is more complex (Pal et al. 2006). The "importance" of proteins, as measured notably by the phenotypic effect of knocking the genes out, predicts only a small fraction of variability. Instead, the strongest predictor of protein evolutionary rates, at least in yeast and *E. coli*, appears to be the level of expression of the corresponding gene (Rocha and Danchin 2004). Other significant factors, with a smaller contribution, include mutation rates, recombination rates, protein tertiary structure, and protein-protein interactions (Pal et al. 2006). In mammals, the relation of protein sequence evolutionary rate with expression level is weaker, and is mostly explained by breadth of expression among tissues (Duret and Mouchiroud 2000; Gu and Su 2007;

Larracuente et al. 2008; Kryuchkova-Mostacci and Robinson-Rechavi 2015), and by expression levels in neural tissues (Gu and Su 2007; Drummond and Wilke 2008; Kryuchkova-Mostacci and Robinson-Rechavi 2015). There is also a correlation in mammals, but not in yeasts, between protein sequence evolutionary rate and changes in expression (Warnefors and Kaessmann 2013).

This variation in mean evolutionary rates reflects differences in purifying selection on protein structure and its capacity to carry out its function. Proteins with different functions are also obviously affected differently by such purifying selection, for two reasons: some gene functions are under stronger selection than others, because they impact phenotype more directly or because they are related to phenotypes which are themselves under stronger selection; and some functions are more directly carried by a specific protein sequence, whereas others less so. For example, histone proteins interact with their whole protein sequence with DNA, thus selection affects all the sequence; and the function of chromatin organisation is fundamental to all cells of an organism, and is under very strong selection. As a result, histones have among the lowest sequence evolutionary rates of any proteins. On the other hand, transcription factors such as the Hox genes are also under strong phenotypic selection, as shown by the conservation of the family (Hoegg and Meyer 2005), its chromosomal organisation and expression patterns, among distant animals (Hrycaj and Wellik 2016). Yet Hox protein sequences, like those of many other transcription factors, are very lowly conserved outside of the DNA-binding domain (Hueber et al. 2010). The strong purifying selection does not seem to act directly on most of the protein sequence. Thus different functional categories of genes are under different selective regimes concerning their protein sequences. An additional selective pressure on protein evolutionary rates is that in some tissues, or for some functions, errors in protein synthesis or protein variants have a higher chance of producing misfolded proteins which are toxic to the cell. This leads to optimization of gene sequence to minimize translation and folding errors, and greater intolerance to some types of mutations (Drummond and Wilke 2008; Drummond and Wilke 2009; Singh et al. 2012).

> Genes rarely act in isolation, but rather as complexes, networks, or pathways. The information on these gene and protein interactions is difficult to measure accurately at a large scale. Metabolic networks or gene regulatory networks typically integrate information from thousands of precise small-scale experiments, only available in a very small number of model species. Metabolic

> networks are especially useful to study the phylogenomics of unicellular organisms, and notably bacteria, where evolution by gene gain (by horizontal transfer) and loss is important, and can be understood as adding or removing nodes from such networks (Pal et al. 2005; Noda-Garcia et al. 2018). Gene regulatory networks are especially attractive because they provide a link between phylogenomics and Evo-Devo (Davidson and Erwin 2006), but robust data at a large scale is rare. Protein-protein interaction networks have been published for several model species, but they still sample the tree of life very sparsely. They have been useful in characterizing differences in evolutionary patterns, e.g., between hub and peripheral proteins (Mintseris and Weng 2005; Wapinski et al. 2007; Presser et al. 2008), but data sampling and quality are so far not sufficient to directly compare homologous proteins and study the evolution of function (Presser et al. 2008).

**Box 3: Network definitions of function**

Protein function also affects sequence evolution through variation in the extent and the mode of positive selection. Continuous positive selection over long evolutionary time has mostly been found on genes involved in sexual selection or immune systems (Obbard et al. 2009; Enard et al. 2016), while episodic positive selection has been found in a wider range of functions (Kosiol et al. 2008; Studer et al. 2008; Barreiro and Quintana-Murci 2010; Daub et al. 2013; Daub et al. 2017; Slodkowicz and Goldman 2019). Positive selection patterns are also affected by expression, with more adaptation in genes expressed in the germ-line (Salvador-Martínez et al. 2018), and of genes expressed post-embryonically rather than embryonically (Liu and Robinson-Rechavi 2018; Coronado-Zamora et al. 2019). Such results are of course dependent on the quality of our positive selection predictions, but they show that to understand adaptation in phylogenomics, we need to take into account gene function.

### 3.2. Duplication and loss: conservative and dynamic functions

The main mechanism by which genes diversify within genomes is duplication. Different molecular mechanisms, such as non-homologous crossover, or transposition, can lead to a DNA region containing one or more genes to be in two or more copies in one haploid genome. Hybridization or abnormal meiosis can lead to polyploidy, in which an individual has extra copies of the whole genome. It is important to keep in mind that these events are mutations. Thus they follow the same dynamics and forces as all mutations. They can rise to fixation in a population or not, under a combination of selection and drift. When polyploidy rises to fixation, and the paralogous copies start diverging, it is often called whole genome duplication (Wolfe 2001). From the perspective of the evolution of gene function, whole genome duplication and small-scale duplication have important differences (Figure 1). A whole genome duplication means that duplication of all genes goes to fixation without any impact of the function of each gene. It also means that each gene is duplicated with its full genomic environment, including promoters and enhancers, and that stoichiometry between all gene products is maintained. Conversely, after small-scale duplication, the fixation of the individual duplicated gene will be affected by selection on that gene's function. And duplicate genes can be unequal "at birth" (Kaessmann et al. 2009), if one copy lacks some regulatory elements due to a partial duplication. In all cases, after fixation, duplicate genes can be retained or not. Duplicates are not retained if one copy suffers a nonsense mutation and becomes a pseudogene, and is then eliminated from the genome. If both copies are kept, they can keep the same function or diverge in function (see below).

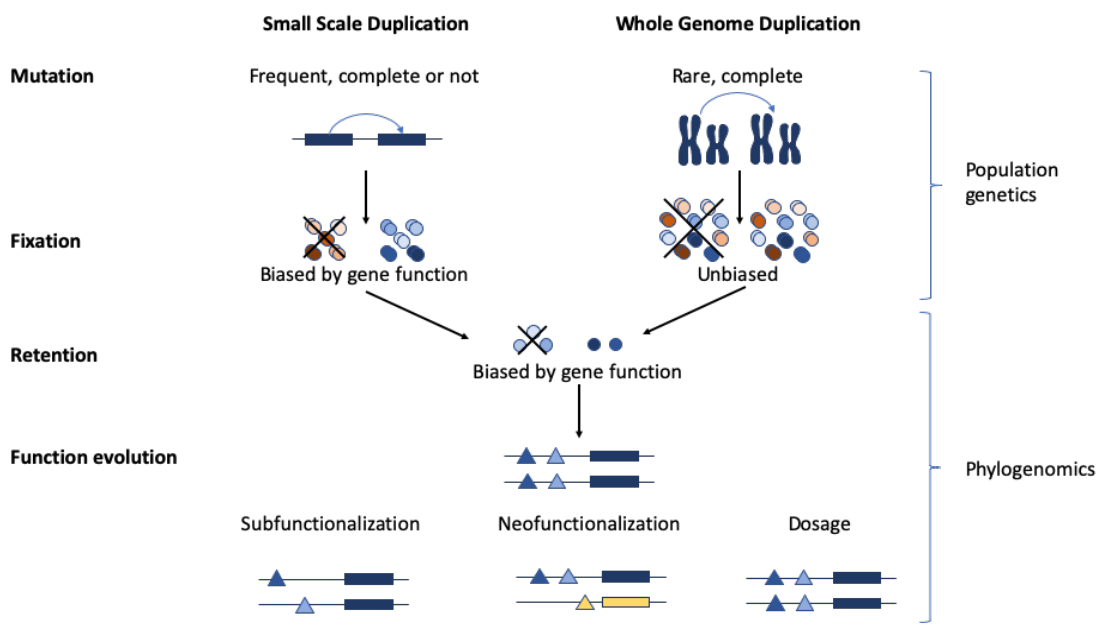

**Figure 1: Dynamics of gene duplication evolution from a functional perspective. In the bottom section of the figure, the triangles represent subfunctions of each gene, for example different regulatory elements.**

As small-scale duplication is much more common (according to some estimates (Lynch and Conery 2000), as common as point mutation), it has the largest impact on overall phylogenomics. The function of genes affects their duplication patterns. Functional biases can be at the mutation level (higher probability of duplicating shorter genes, or genes expressed in the germinal line), as well as fixation and retention (Figure 1). Some functional categories tend to duplicate and be lost from genomes (i.e., turn-over) much more. Other functional categories are very conservative, and are mostly found as 1-to-1 orthologs between species. Some of the same functional categories which evolve rapidly at the sequence level also have a large turn-over of gene copy number (Heger and Ponting 2007; Ponting 2008), notably immune defence and host evasion, and reproduction. These functions thus evolve rapidly both by amino acid substitutions and by duplication and loss of genes, allowing rapid adaptation, typically within arms-race contexts. Another functional class with abundant turn-over is metabolism genes (Demuth and Hahn 2009), whereas these genes tend to evolve conservatively in protein sequence. Variation in copy number of metabolism genes can either contribute to the functional diversity of metabolic pathways, or to changes in dosage of metabolism proteins. Whatever the patterns of duplication, some functions seem more resistant to gene loss (Albalat and Cañestro 2016), probably due to low dispensability of the specific function of genes in those categories.

Observed patterns of gene duplication are in great part due to variations in the selection pressure that drives paralog retention or loss after the duplication event itself. From this point of view, there are important differences between whole genome duplications and small-scale duplications. All genes are duplicated in a genome duplication, and there are no issues of stoichiometry nor of missing regulatory regions for some duplicate copies. Thus the impact of gene function on retention is not biased by other processes. Studies have found long term retention of 10-20% of duplicate genes after whole genome duplication (Wolfe 2001; Jaillon et al. 2004; Nakatani et al. 2007; Putnam et al. 2008). There is strong evidence that this loss of duplicates is non-random, and thus enriches genomes in specific classes of genes (Davis and Petrov 2004; Brunet et al. 2006; Roux and Robinson-Rechavi 2008; Makino et al. 2009; Gout et al. 2010; Makino and McLysaght 2012). In vertebrates, for example, this biased retention seems largely driven by selection against detrimental mutations of genes. This leads to a pattern of retention of genes whose variants have a higher chance of being toxic (see selection against protein misfolding above), such as those involved in diseases (Singh et al. 2014), and of genes highly expressed in the nervous system (Roux et al. 2017).

While there are general trends in gene turn-over for broad categories, many specific gene family expansions or losses are lineage-specific (Lespinet et al. 2002). There are biases in gene "duplicability" which affect the small-scale duplications, which lead to such expansions, and unlike for whole genome duplication, all steps can be biased, from the duplication mutation itself to fixation, and to retention. As an example of mutation bias, there are more retrogenes from genes expressed in testis in mammals (Kaessmann et al. 2009). Fixation bias appears to go in the opposite direction for small-scale duplicate genes than for genome duplication, with genes under strong purifying selection being eliminated before fixation as paralogs (Rice and McLysaght 2017; Roux et al. 2017). While these mechanisms are mostly due to the varying strength of purifying selection, gene family expansions of some functional categories appear to be good candidates for adaptation. For example, olfactory receptors have repeatedly expanded in lineages such as fishes, mammals, or ants (Hussain et al. 2009; Niimura et al. 2014; McKenzie and Kronauer 2018).

Gene function affects every step of the evolutionary dynamics of duplication, and ignoring the biases in generation, fixation, and retention of paralogs can lead to wrong inferences (Davis and Petrov 2004; Studer and Robinson-Rechavi 2009). This is a more general lesson: to study the evolution of gene function we should always control for the ways in which function can impact evolution upstream of the changes we want to study.

## 4. How does gene function evolve?

In addition to the impact of function on gene evolution, the function of genes itself evolves. This is in principle the most interesting aspect of the phylogenomics of function. Yet it is poorly known because this is where the difficulties in defining gene function are the most disturbing. The impact of function on gene evolution is evident through large differences between broad categories. Low granularity of functional classification is sufficient to show that immune system genes evolve under stronger positive selection, or that genes expressed in the nervous system are more often kept in several copies after genome duplication. But the evolution of gene function very rarely consists in shifts between these broad categories. Indeed, the success of gene and protein domain annotation by homology (Jiang et al. 2016) testifies to the rarity of radical shifts in function during gene evolution. Such shifts do occur, most dramatically illustrated by crystallins in tetrapod eyes (reviewed in Graur 2016). For example in rabbits cystallin λ is a paralog of a dehydrogenase, and in frogs crystallin ρ is a paralog of a reductase. Sometimes the same protein carries both an enzymatic function and the crystallin function, known as "moonlighting proteins" (Jeffery 2018), for example crystallin ε in crocodiles and ducks which is also a lactate dehydrogenase. Such cases remain rare as far as we know. Transcription factors remain transcription factors, but change subtly their specificity, affinity, or timing of expression. Membrane receptors remain receptors, but evolve different co-factors, or shift affinity for different ligands. Thus the study of the evolution of gene function is limited by our capacity to determine function of homologous genes both accurately and in an unbiased manner.

### 4.1. Evolution of gene expression

Gene expression patterns have consistently been a key feature used to characterize the evolution of function. Expression can be measured easily in diverse species, it is immediately comparable between genes that are otherwise very different (unlike, e.g., comparing the activity of a transcription factor and of an enzyme), and it lends itself well to modelling. With modern techniques it also lends itself well to large-scale studies, such as RNA-seq, including in non-model organisms.

A notable example is the original model of sub-functionalization by Duplication-Degeneration-Complementation (DDC), which was derived from small-scale observations of gene expression in fish and mammalian development (Force et al. 1999) (Figure 1, section Function evolution). While it is clear that gene function can change in evolution without change in expression pattern, a change in expression pattern between homologs can be interpreted as indicating that at least some aspect of the function has changed. In the DDC sub-functionalization model applied to expression patterns, paralogs evolve from an ancestral gene which has several domains of expression, and by losing different domains of expression in each paralog, end up recapitulating between them the ancestral pattern which neither covers entirely alone. These domains of expression can be anatomical domains (tissues, organs, cell types), timing of expression (e.g., over development), or any other aspect of expression (e.g., reaction to extrinsic signals, or sex bias). Thus for example after duplication of a gene expressed in the pectoral appendage bud and in the hindbrain in fish embryos, one paralog might conserve expression in the pectoral appendage bud, and the other in the hindbrain (this is the *eng1a/b* example used in (Force et al. 1999)).

There have been many attempts to test this model, and while results have been mixed for the specific DDC model, they show that expression patterns, combined or not with information on expression levels, can be successfully used to study at least some aspects of gene function. For example, comparisons of expression patterns of genes in teleost fish after genome duplication to non-duplicate gar outgroup orthologs provided support for sub-functionalization, with typical patterns of each paralog expressed in different tissues, and the non-duplicated ortholog expressed in both (Braasch et al. 2016). The same study showed quantitative subfunctionalization, with the expression levels of two paralogs recapitulating the level of non-duplicated genes. Conversely, a study of expression of genes duplicated in the salmonid genome duplication found a dominant pattern of neo-functionalization, with one conserved paralog and one diverged: the former expressed in the same pattern as the non-duplicated ortholog, the latter expressed in different organs (Lien et al. 2016). A re-analysis of both studies indicates support for asymmetric evolution, but is not conclusive on sub- vs. neo-functionalization (Sandve et al. 2018).

## 4.2. The Ortholog Conjecture and the difficulty of assessing function evolution

Phylogenomics comparisons of function in the absence of duplication have been complicated, because the problems discussed in the first section of this chapter complicate defining a null expectation. Conservation of function can be measured in some cases (e.g. of expression among mammals in (Brawand et al. 2011; Piasecka, Kutalik, et al. 2012)), but distinguishing functional change from errors in the data and analysis is extremely difficult. A case study, which nicely illustrates the difficulties of studying gene function evolution at a phylogenomic scale, is the question of the "ortholog conjecture".

The ortholog conjecture is the hypothesis that orthologous genes have mostly conserved function, or that their function diverges very slowly during evolution, whereas paralogous genes have mostly different functions, or that their function diverges very rapidly during evolution (Figure 2). While it was a foundational hypothesis of phylogenomics (Eisen 1998), it has only started being tested systematically (and named) in the last 10 years (Studer and Robinson-Rechavi 2009; Nehrt et al. 2011). The ortholog conjecture has been surprisingly difficult to confirm or infirm robustly, using diverse datasets and definitions of gene function.

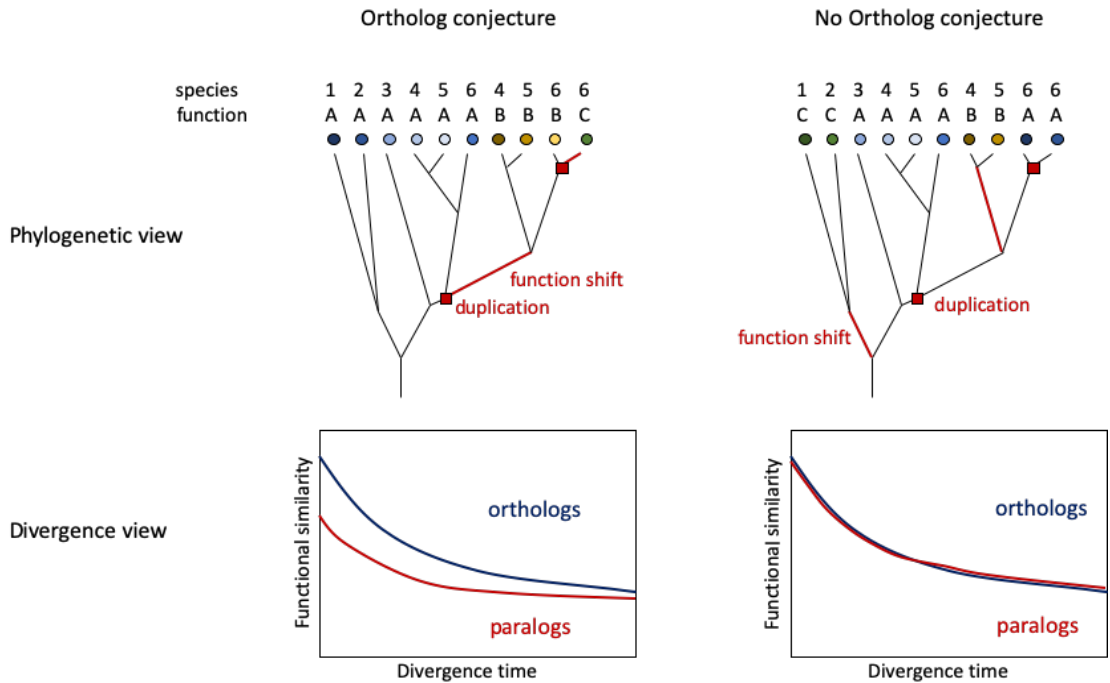

**Figure 2: Schematic expectations of function evolution between orthologs and paralogs.** Left, expectations under the ortholog conjecture, right, expectations if this conjecture is not supported (under a naive null of random functional changes during gene evolution). Phylogenetic view: gene tree with gene duplications indicated by red squares and functional shifts by red branches; the coloured circles are homologous genes, with the colour according to similarity of function; above, species identity (notice that following duplication, some species are represented several times in the tree) and functional classification as might be captured e.g. by the Gene Ontology. Notice that paralogs within one species might have different functions even if the ortholog conjecture is wrong, e.g. the paralogs in species 4 and 5. Divergence view: expectation of functional divergence between pairs of orthologs and of paralogs; in all cases, functional similarity is expected to decrease with evolutionary time, but paralogs are expected to diverge more and faster than orthologs under the ortholog conjecture.

Two of the first studies on the ortholog conjecture used the Gene Ontology to define functional divergence in proportion to the difference in GO annotations between genes (Nehrt et al. 2011; Altenhoff et al. 2012). Both studies took into account the ontology graph, i.e. that a hydrolase is necessarily also an enzyme, but obtained opposing results. The second study showed that paralogs in the same species tend to be studied by the same research groups, leading to similar experiments and annotations, whereas orthologs tend to be studied by different groups, leading to different experiments and annotations (see Table 1). This biases GO comparisons towards apparently more similar functional annotations between paralogs, whereas correcting for it shows more similar functional annotations between orthologs, although the effect is small (Altenhoff et al. 2012). In an unusual move, the leaders of the GO consortium published a short paper explaining why GO annotations could not be used to study evolutionary patterns of function (Thomas et al. 2012). Finally, the evolution of GO annotations over time makes any evolutionary interpretation very difficult (Chen and Zhang 2012).

Most subsequent studies of the ortholog conjecture have focused on gene expression, for the same reasons as in other studies of gene function and evolution. Using correlations of expression levels within and between species, different studies again reached different conclusions depending on methods. Microarray data comparison was not consistent with the ortholog conjecture (Nehrt et al. 2011), but this might be due to differences in microarrays between species (Liao and Zhang 2006; Chen and Zhang 2012). Comparing expression levels from RNA-seq provides support for the ortholog conjecture (Chen and Zhang 2012; Rogozin et al. 2014), although the effect size is weak and depends on the correlation method used. To avoid these issues with comparing expression levels between species, we summarized expression across tissues by the measure of "tissue-specificity", and found that it is well conserved between orthologs, different between paralogs, and diverges with time, as expected from the ortholog conjecture, and with large effect size of the difference between orthologs and paralogs (Kryuchkova-Mostacci and Robinson-Rechavi 2016). But a reanalysis pointed out that pairwise comparisons are biased when studying evolutionary changes. Using a phylogenetic framework on the same tissue-specificity data, the support for the ortholog conjecture disappears (Dunn et al. 2018).

These conflicting results show that even for a very well defined question (do paralogs diverge more than orthologs of the same age?), it is very difficult to study rigorously the evolution of gene function on a genomic scale.

## 5. Conclusions

The fundamental reason that we are interested in gene evolution in phylogenomics, as opposed to the evolution of random sequences of DNA, is that they carry functions, which relate the genome to the phenotype and organismal fitness. Thus we would like both to study the evolution of genes in the context of their function, allowing us to study the evolution of functional units, and to study how the function of the genes themselves evolves. On the first aim, research in the last 20 years has provided us with a view of how purifying and adaptive selection affect functional units, but limited to a very broad definition of these units: highly expressed genes, proteins central in interaction networks, potentially toxic proteins, etc. On the second aim, this lack of precision proves to be extremely limiting, and we still know surprisingly little about how gene function evolves. The difficulties in testing the "ortholog conjecture" illustrate this: if we are unable to verify such a basic assumption of our field, it seems difficult to discover new patterns until we have further improved our data and methods. Finally, the study of molecular evolution and function is in the same boat as much of genomics, suffering from too much vagueness around the notion of function (Doolittle 2018).